\documentclass[%
 reprint,
superscriptaddress,
 amsmath,amssymb,
 aps,
]{revtex4-2}

\usepackage{graphicx}
\usepackage{dcolumn}
\usepackage{bm}
\usepackage{multirow}
\usepackage{hyperref}

\begin{document}


\title{Spontaneous polarization in NaNbO$_{3}$ film on NdGaO$_{3}$ and DyScO$_{3}$ substrates}

\author{Kisung Kang}
 \email{kang@fhi-berlin.mpg.de}
\affiliation{ 
The NOMAD Laboratory at the FHI of the Max-Planck-Gesellschaft and IRIS-Adlershof of the Humboldt-Universit\"{a}t zu Berlin, Faradayweg 4-6, 14195 Berlin, Germany
}

\author{Saud Bin Anooz}
\affiliation{ 
Leibniz-Institut f\"{u}r Kristallz\"{u}chtung (IKZ), Max-Born-Str. 2, 12489 Berlin, Germany
}

\author{Jutta Schwarzkopf}
\affiliation{
Leibniz-Institut f\"{u}r Kristallz\"{u}chtung (IKZ), Max-Born-Str. 2, 12489 Berlin, Germany
}

\author{Christian Carbogno}
\affiliation{
The NOMAD Laboratory at the FHI of the Max-Planck-Gesellschaft and IRIS-Adlershof of the Humboldt-Universit\"{a}t zu Berlin, Faradayweg 4-6, 14195 Berlin, Germany
}

\date{\today}

\begin{abstract}
Pure NaNbO$_{3}$ is an antiferroelectric material at room temperature that irreversibly transforms to a ferroelectric polar state when subjected to an external electrical field or lattice strain.
Experimentally, it has been observed that NaNbO$_{3}$ films grown on NdGaO$_{3}$ exhibit an electrical polarization along the [001]$_{\mathrm{PC}}$ direction, whereas films on DyScO$_{3}$ substrates exhibit a polarization along the [011]$_{\mathrm{PC}}$ direction.
These effects have been attributed to the realization of different lattice symmetries in the films due to the incorporation of lattice strain imposed by the use of oxide substrates with different lattice parameters.
However, the underlying atomistic mechanisms of the resulting phase symmetry in the films are hardly clear, given that NaNbO$_{3}$ features a diverse and complex phase diagram.
In turn, these also impede a straightforward tailoring and optimization of the resulting macroscopic properties on different substrates.
To clarify this issue, we perform all-electron first-principles calculations for several potential NaNbO$_{3}$ polymorphs under stress and strain.
The computed properties, including the ferroelectric polarization, reveal that an orthorhombic $Pmc2_{1}$ phase is realized on NdGaO$_{3}$ substrates since this is the only phase with an out-of-plane polarization under a compressive strain.
Conversely, the monoclinic $Pm$ phase is consistent for the samples grown on DyScO$_{3}$ substrate, since this phase exhibits a spontaneous in-plane polarization along [011]$_{\mathrm{PC}}$ under compressive strain.
\end{abstract}

\maketitle


\section{\label{sec:intro}Introduction}

The lead-free perovskite sodium niobate (NaNbO$_{3}$) has been widely characterized and investigated in literature due to its favorable properties,~e.g.,~low coercive fields\cite{Yang:2022, Luo:2023}, a high saturated polarization~\cite{Yang:2022, Chen:2022, Luo:2023}, a high energy density\cite{Chen:2022, Sun:2021}, ultrafast charge/discharge rates~\cite{Jiang:2021, Fan:2019}, and piezoelectric properties~\cite{Reznitchenko:2001}. 
These properties, which are attributed to the existence of ferroelectric (FE)\cite{Arioka:2010} and antiferroelectric (AFE)\cite{Yang:2022, Jiang:2021, Fan:2019, Luo:2023} phases of NaNbO$_{3}$, result in a high potential in a wide range of applications, including  photocatalysis\cite{Li:2008, Katsumata:2009, Yang:2019}, energy storage\cite{Qi:2019, Pan:2017}, and memory devices\cite{Lingwal:2003}. 
For this purpose, strain engineering\cite{Schwarzkopf:2012, Schmidbauer:2014, Sellmann:2014, BinAnooz:2015, BinAnooz:2022, Guimaraes:2022} has been shown as a viable route to further tailor and 
optimize the properties of NaNbO$_{3}$.

In this regard, the structural phase richness~\cite{Megaw:1974, Ringgaard:2005} of NaNbO$_3$ is a double-edged sword.
On the one hand, it is key for tunable and favorable material properties; on the other hand, it massively complicates modeling and understanding the underlying physics.
In bulk NaNbO$_{3}$, seven different bulk phases at various temperatures and normal pressure conditions have been reported.\cite{Megaw:1974}
Notably, a distinctive ferroelectric-antiferroelectric phase transition from $R3c$ to $Pbcm$ at 173\,K has been observed through Raman scattering\cite{Shen:1998} and neutron diffraction.\cite{Mishra:2007}
Given the relatively small energy difference between FE and AFE phases\cite{Shimizu:2015}, external stimuli,~e.g.,~the application of an electrical field\cite{Ulinzheyev:1990, Zhang:2020, Zhang:2021} or of mechanical stress\cite{Schwarzkopf:2012, Schmidbauer:2014, Sellmann:2014, BinAnooz:2015, BinAnooz:2022, Guimaraes:2022, Koruza:2017, Hu:2023}, can alter the relative stability of the individual FE and AFE phases.

An illustrative example is the heteroepitaxial growth of NaNbO$_{3}$ thin films using metalorganic vapor-phase epitaxy (MOVPE)\cite{Schwarzkopf:2012}.
Due to the lattice mismatch between film and substrate\cite{Schwarzkopf:2012, Schmidbauer:2014, Sellmann:2014, BinAnooz:2015, BinAnooz:2022, Guimaraes:2022}, elastic strain is induced.
By using different perovskite substrates,~e.g.,~NdGaO$_{3}$, SrTiO$_{3}$, DyScO$_{3}$, TbScO$_{3}$, or GdScO$_{3}$, it becomes possible to manipulate the magnitude of the in-plane elastic strain from -0.007 (compressive) to 0.014 (tensile).\cite{Schwarzkopf:2012}
This variation of strain results in a stabilization of different polymorphs or even of structures not known to be realized in the bulk,~e.g.,~a monoclinic structure\cite{Schwarzkopf:2012}.
In turn, different crystal symmetries with varying ferroelectric polarizations are observed when growing  NaNbO$_{3}$ on different substrates.
However, identifying the precise structural phase from the numerous potential structures remains challenging.

To obtain insights into the stability range and properties of the different NaNbO$_3$ polymorphs, theoretical investigations have been proven helpful in the past.
For instance, Di\'{e}guez and his colleagues utilized the first-principles energy parameterization method developed by King-Smith and Vanderbilt\cite{King-Smith:1994} to examine the influence of in-plane strain on NaNbO$_{3}$ and to predict the emergence of a monoclinic structure with $Cm$ space group.\cite{Dieguez:2005}
More recently, a density functional theory study by Patel and collaborators proposed two new low-energy orthorhombic phases ($Pmc2_{1}$ and $Pca2_{1}$) and a ferroelectric monoclinic phase $Cc$, all of which are stabilized by tensile strain.\cite{Patel:2021}
Combining and comparing such first-principles investigations with experimental data is hence a viable route to clarify which specific phases manifest in real film samples on different substrates.

In this spirit, we investigate the phases of epitaxial NaNbO$_{3}$ films on NdGaO$_{3}$ and DyScO$_{3}$ substrates through density-functional theory calculations and compare our findings with experimental evidence\cite{Schwarzkopf:2012, Schwarzkopf:2014, Schmidbauer:2014, Sellmann:2014, BinAnooz:2015, BinAnooz:2022, Guimaraes:2022}. 
We study ten different phases of NaNbO$_{3}$ and meticulously compare the calculated properties with measured data\cite{Schwarzkopf:2012, Schwarzkopf:2014, Schmidbauer:2014, Sellmann:2014, BinAnooz:2015, BinAnooz:2022, Guimaraes:2022}.
While structural information such as lattice constant and electronic properties such as band gaps provide some insights that allow us to rule out some candidate phases, they alone are insufficient for identifying the structure of the film.
However, additional calculations of the electrical polarization, its alignment, and its dependence on strain allowed us to single out the phases realized on NdGaO$_{3}$ and DyScO$_{3}$ substrates.

\section{\label{sec:comp}Computational Details}
In the following, we present density-functional theory (DFT) calculations performed with the all-electron code \texttt{FHI-aims}\cite{Blum:2009, Knuth:2015} utilizing numeric atom-centered orbitals as a basis.
Exchange and correlation effects are described within either the generalized gradient approximation using the PBEsol functional\cite{Perdew:2008} or the hybrid functional proposed by Heyd, Scuseria, and Ernzerhof (HSE06)\cite{Krukau:2006} using ``tight'' and ``intermediate'' numerical and basis set defaults, respectively.
To achieve convergence in total energy below 0.05\,meV/atom, the Brillouin zone of the 
$Pm$-$3m$, $P4/mbm$, $R3c$, $Cc$, $Pm$, $Cmcm$, $Pca2_{1}$, $Pbcm$, $Pnma$, and $Pmc2_{1}$ structures sampled with Monkhorst-Pack\cite{Monkhorst:1976} grids of size  
$8\times8\times8$    ($6\times6\times6$), 
$8\times8\times10$   ($8\times8\times10$),    
$9\times9\times9$    ($9\times9\times9$), 
$12\times12\times7$  ($7\times7\times4$),   
$10\times10\times10$ ($10\times10\times10$),  
$5\times5\times5$,   ($4\times4\times4$), 
$6\times6\times2$,   ($9\times9\times3$), 
$8\times8\times3$,   ($8\times8\times3$), 
$10\times10\times7$  ($7\times7\times5$),   
and $12\times12\times8$  ($7\times7\times5$)      
when using the PBEsol~(HSE06) functional.
All electronic band structures are calculated perturbatively, on top of self-consistent calculations with a twice as dense {\bf k}-point grid, and by additionally including spin-orbit coupling~\cite{Huhn:2017}.
The electrical polarization~\cite{Resta:1992} of NaNbO$_{3}$ is evaluated in a Berry-phase-type approach~\cite{King-Smith:1993} using the cubic $Pm$-$3m$ phase as a reference with vanishing polarization\cite{Spaldin:2012}.
For this purpose, we employ a denser {\bf k}-point sampling scheme for all distinct phases.
For example, when using $h \times k \times l$ {\bf k}-points for relaxation, we sample $3h$ {\bf k}-points along the reciprocal-space line integral and average over $2k \times 2l$ of these paths to determine the polarization along the $h$ direction.

\section{\label{sec:res}Results and Discussion}

\begin{table}
\caption{\label{tab:struc} Measured pseudocubic (PC) lattice parameters of NaNbO$_{3}$ films on NdGaO$_{3}$ and DyScO$_{3}$ substrates~(Exp.)\cite{Schmidbauer:2014, BinAnooz:2015} compared to computed lattice parameters (PBEsol and HSE06) for the bulk orthorhombic $Pmc2_{1}$ and monoclinic $Pm$ phase. 
The percentage difference between the measured and calculated lattice parameters is presented.
In the PC cell notation, all lattice parameters are given with respect to a non-primitive, five-atom unit cell, which is similar to the $Pm$-$3m$ primitive cell, but allows for tiltings and distortions\cite{Vailionis:2011}.}

\begin{tabular}{ccccccc}
\\
\multicolumn{7}{l}{NaNbO$_{3}$/NdGaO$_{3}$ ($Pmc2_{1}$)}\\
\hline\hline
& $\mathbf{a}_{\mathrm{PC}}$ & $\mathbf{b}_{\mathrm{PC}}$ & $\mathbf{c}_{\mathrm{PC}}$ & $\mathbf{\alpha}_{\mathrm{PC}}$ & $\mathbf{\beta}_{\mathrm{PC}}$ & $\mathbf{\gamma}_{\mathrm{PC}}$ \\
\hline
Exp.\cite{Schmidbauer:2014, BinAnooz:2015} & 3.863~\AA & 3.854~\AA & 3.984~\AA & & & \\
PBEsol & 3.869~\AA & 3.913~\AA & 3.913~\AA & 89.4$^{\circ}$ & 90$^{\circ}$ & 90$^{\circ}$ \\
Diff. & 0.2~\% & 1.5~\% & -1.8~\% & & & \\
HSE06 & 3.867~\AA & 3.931~\AA & 3.931~\AA & 89.4$^{\circ}$ & 90$^{\circ}$ & 90$^{\circ}$\\
Diff. & 0.1~\% & 2.0~\% & -1.3~\% & & & \\
\hline\hline
\\
\multicolumn{7}{l}{NaNbO$_{3}$/DyScO$_{3}$ ($Pm$)}\\
\hline\hline
& $\mathbf{a}_{\mathrm{PC}}$ & $\mathbf{b}_{\mathrm{PC}}$ & $\mathbf{c}_{\mathrm{PC}}$ & $\mathbf{\alpha}_{\mathrm{PC}}$ & $\mathbf{\beta}_{\mathrm{PC}}$ & $\mathbf{\gamma}_{\mathrm{PC}}$ \\
\hline
Exp.\cite{Schmidbauer:2014, BinAnooz:2015} & 3.849~\AA & 3.947~\AA & 3.952~\AA & 89.5$^{\circ}$ & 90$^{\circ}$ & 90$^{\circ}$ \\
PBEsol & 3.908~\AA & 3.978~\AA & 3.979~\AA & 89.4$^{\circ}$ & 90$^{\circ}$ & 90$^{\circ}$ \\
Diff. & 1.5~\% & 0.8~\% & 0.7~\% & -0.1~\% & 0.0~\% & 0.0~\% \\
HSE06 & 3.902~\AA & 3.983~\AA & 3.984~\AA & 89.3$^{\circ}$ & 90$^{\circ}$ & 90$^{\circ}$\\
Diff. & 1.4~\% & 0.9~\% & 0.8~\% & -0.2~\% & 0.0~\% & 0.0~\% \\
\hline\hline
\end{tabular}

\end{table}

Several experimental measurements provide evidence and hints that help to pinpoint which polymorph is realized under strain when NaNbO$_{3}$ films are grown on NdGaO$_{3}$\cite{Schwarzkopf:2012, Schwarzkopf:2014, Sellmann:2014, BinAnooz:2015, BinAnooz:2022} or DyScO$_{3}$\cite{Schwarzkopf:2012, Schwarzkopf:2014, Schmidbauer:2014, Sellmann:2014, BinAnooz:2015, BinAnooz:2022, Guimaraes:2022}.
These experimental findings, which help guide our computational investigation in identifying the phases of NaNbO$_{3}$ films, include data regarding the lattice system, the band gap, and the electrical polarization.
High-resolution X-ray diffraction (HRXRD) and grazing incidence X-ray diffraction (GIXRD) measurements reveal that NaNbO$_{3}$ films on NdGaO$_{3}$ and DyScO$_{3}$ substrates have orthorhombic and monoclinic lattice systems, respectively\cite{Schwarzkopf:2014, BinAnooz:2015}.
Detailed lattice parameters for these two film samples obtained from the experiments\cite{Schmidbauer:2014, BinAnooz:2015} are listed in Tab.~\ref{tab:struc}.
The electronic band gap energy of NaNbO$_{3}$/NdGaO$_{3}$, determined optically, measures about 3.8 eV\cite{BinAnooz:2022}.
On the other hand, NaNbO$_{3}$/DyScO$_{3}$ samples display band gaps of 3.9 eV\cite{BinAnooz:2022}.
Furthermore, both films exhibit spontaneous electrical polarization, as observed through vertical and lateral piezoresponse force microscope (PFM).
In case of NaNbO$_{3}$/NdGaO$_{3}$, vertical PFM measurments reveal a pronounced out-of-plane polarization signal along [001]$_{\mathrm{PC}}$ \cite{Schwarzkopf:2012, Schwarzkopf:2014}.
For NaNbO$_{3}$/DyScO$_{3}$, lateral PFM images display an in-plane polarization signal along [011]$_{\mathrm{PC}}$, characterized by two alternating ferroelectric 90$^{\circ}$ domains appearing as a stripe pattern\cite{Schwarzkopf:2012, Schmidbauer:2014, Sellmann:2014, Guimaraes:2022}.

\begin{figure*}
\includegraphics[width=1.99\columnwidth]{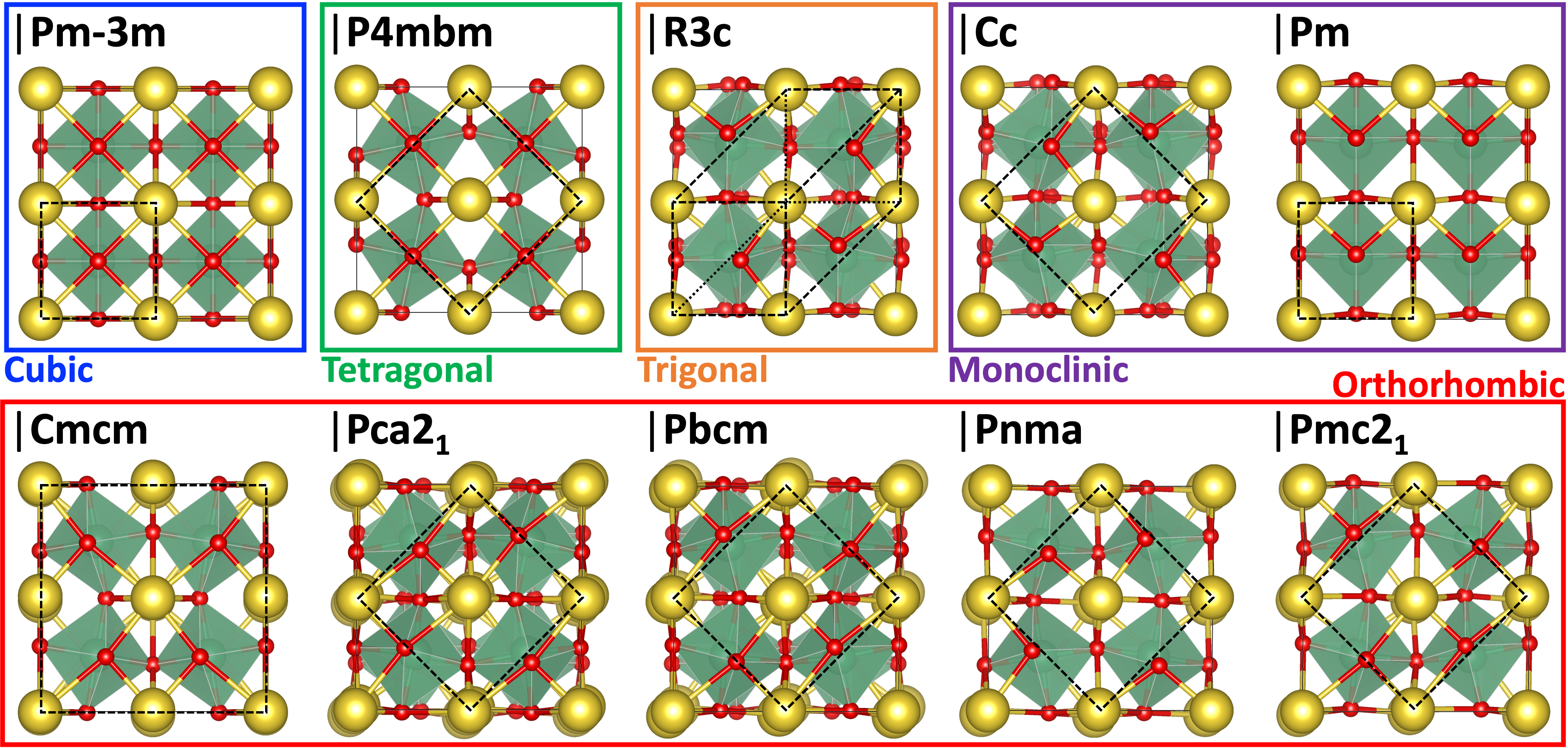}
\caption{\label{fig:struc} All the different phases of NaNbO$_{3}$ investigated using the DFT method. Yellow, green, and red spheres represent Na, Nb, and O elements, respectively. The green polyhedral is centered around the Nb atom and includes its neighboring O atoms. The black dashed lines outline the unit cell of each phase. For electrical polarization calculations, the $Pm$-$3m$ structure is used as the reference state.}
\end{figure*}

\begin{figure}
\includegraphics[width=0.80\columnwidth]{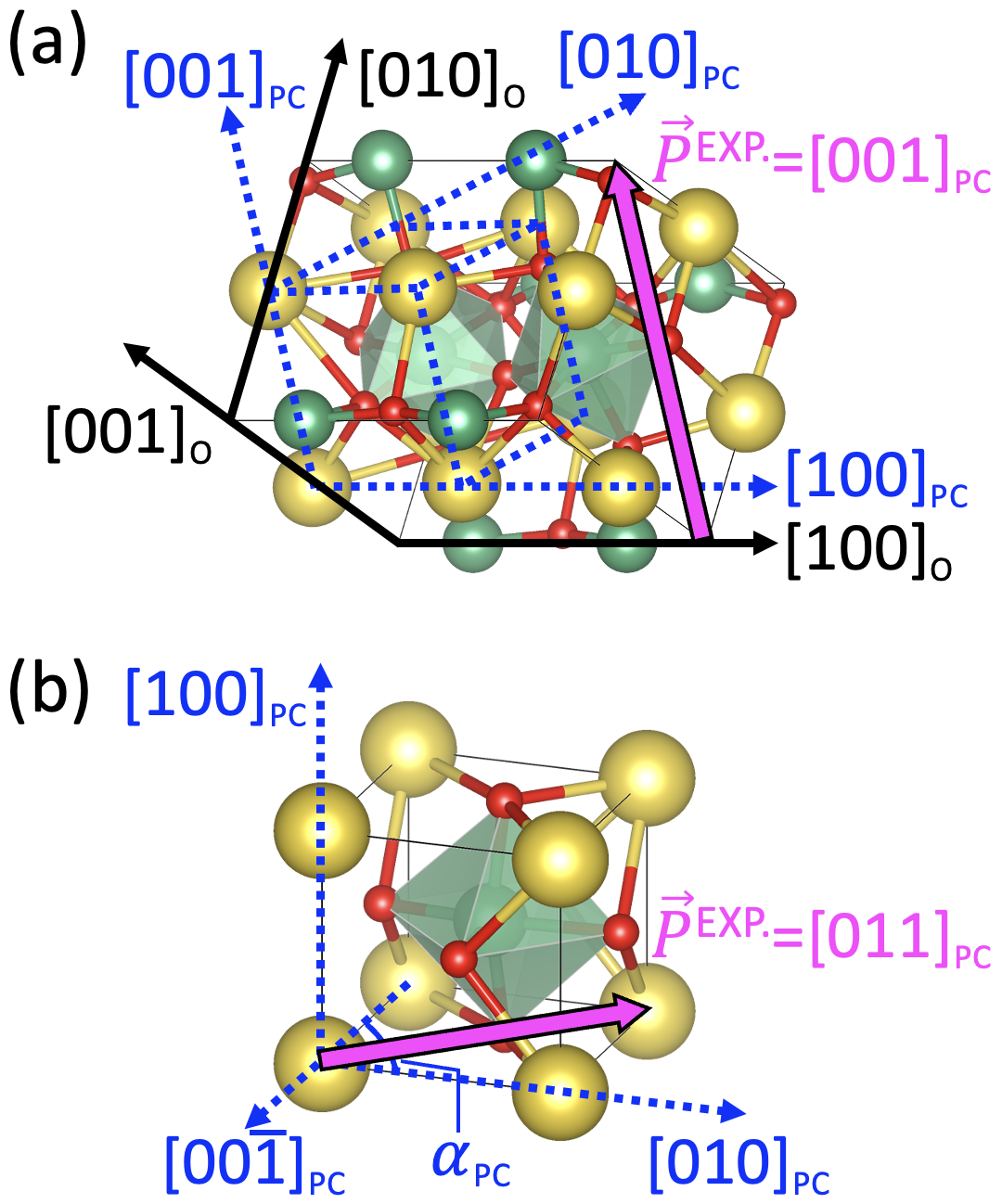}
\caption{\label{fig:axis} (a) An atomic structure of $Pmc2_{1}$ phase annotated with orthorhombic axes, [hkl]$_{\mathrm{O}}$, and pseudocubic (PC) axes, [h'k'l']$_{\mathrm{PC}}$ and (b) an atomic structure of $Pm$ phase annotated with pseudocubic (PC) axes which is identical with monoclinic axes. Magenta arrows show the direction of electrical polarization measured in experiments\cite{Schwarzkopf:2012, Schwarzkopf:2014, Schmidbauer:2014, Sellmann:2014, Guimaraes:2022}.}
\end{figure}

As a starting point in the computational approach, we begin by examining the lattice system of NaNbO$_{3}$.
We investigate a total of ten phases of NaNbO$_{3}$, which include those reported in previous studies of NaNbO$_{3}$~($Pm$-$3m$, $P4/mbm$, $R3c$, $Cc$, $Cmcm$, $Pca2_{1}$, $Pbcm$, $Pnma$, and $Pmc2_{1}$)\cite{Megaw:1974, Patel:2021} as well as a
 monoclinic structure that has been reported for other perovskite materials in the Materials Project database\cite{MP:2020, Jain:2013} and that could hence be potentially realized in  NaNbO$_{3}$.
As illustrated in Fig.~\ref{fig:struc}, these phases include cubic $Pm$-$3m$, tetragonal $P4/mbm$, and trigonal $R3c$   NaNbO$_{3}$ as bulk reference points, 
two monoclinic structures $Cc$ and $Pm$, as well as five orthorhombic structures $Cmcm$, $Pca2_{1}$, $Pbcm$, $Pnma$, and $Pmc2_{1}$. We note that the film samples exhibit a strained structure, 
so that a comparison of the lattice parameters experimentally determined for the strained film with those computed for the unstrained bulk phases does not allow any conclusive insights. 
For example, the NaNbO$_{3}$ film on the NdGaO$_{3}$ substrate exhibits an orthorhombic structure, as shown in Fig.~\ref{fig:axis} (a).
When converting the orthorhombic (O) axes to pseudocubic (PC) axes, $\mathbf{b}_{\mathrm{PC}}$ and $\mathbf{c}_{\mathrm{PC}}$ should be identical because they are half of the diagonal on $\mathbf{bc}$ plane.
Thus, the difference observed between the measured $\mathbf{b}_{\mathrm{PC}}$ and $\mathbf{c}_{\mathrm{PC}}$ in NaNbO$_{3}$ film on the NdGaO$_{3}$ substrate indicates the presence of strain within this film structure.

Nonetheless, it is instructive to compare the measured lattice parameters with those calculated for the orthorhombic and monoclinic candidates, which are listed in Tab.~\ref{tab:struc} and Tab.~S2 in the supplementary material\cite{Supplement}.
For instance, the orthorhombic $Cmcm$ and the monoclinic $Cc$ structure can be immediately sorted out from the list of candidates due to their unitcell shape (See Fig.~\ref{fig:struc}).
Since a $\sqrt{2}\times\sqrt{2}\times{2}$ PC supercell, cf.~Fig.~\ref{fig:axis}(a), is realized for NaNbO$_{3}$/NdGaO$_{3}$ films, the $Cmcm$ structure, which features a $2\times2\times2$ PC supercell as its unitcell, can be ruled out for the NdGaO$_{3}$ substrate. Similarly, the $Cc$ structure, which possesses a $\sqrt{2}\times\sqrt{2}\times2$ PC supercell as its unitcell, can be ruled out for the DyScO$_{3}$ substrate, 
on which NaNbO$_{3}$ is known to grow with a single PC unitcell, see Fig.~\ref{fig:axis}(b).
More quantitatively, however, differentiating between crystal structures within the same lattice systems is challenging, since only subtle variations associated with slightly different tiltings of the octahedra are observed.
For example, all remaining structures differ in their lattice parameters at most by 2.0\% from the measured one. Besides the role of strain, such slight variations can also
be caused by deficiencies in the exchange-correlation functional used in the calculations, as showcased by comparing measured and computed bulk lattice constants for the reference structures ($Pm$-$3m$, $P4/mbm$, and $R3c$) in Tab. S1\cite{Supplement}. Consequently, relying solely on lattice parameters is insufficient to identify the exact phase of the NaNbO$_{3}$ film, although this information about the lattice system does help narrow down the list of candidates.

\begin{figure}
\includegraphics[width=0.98\columnwidth]{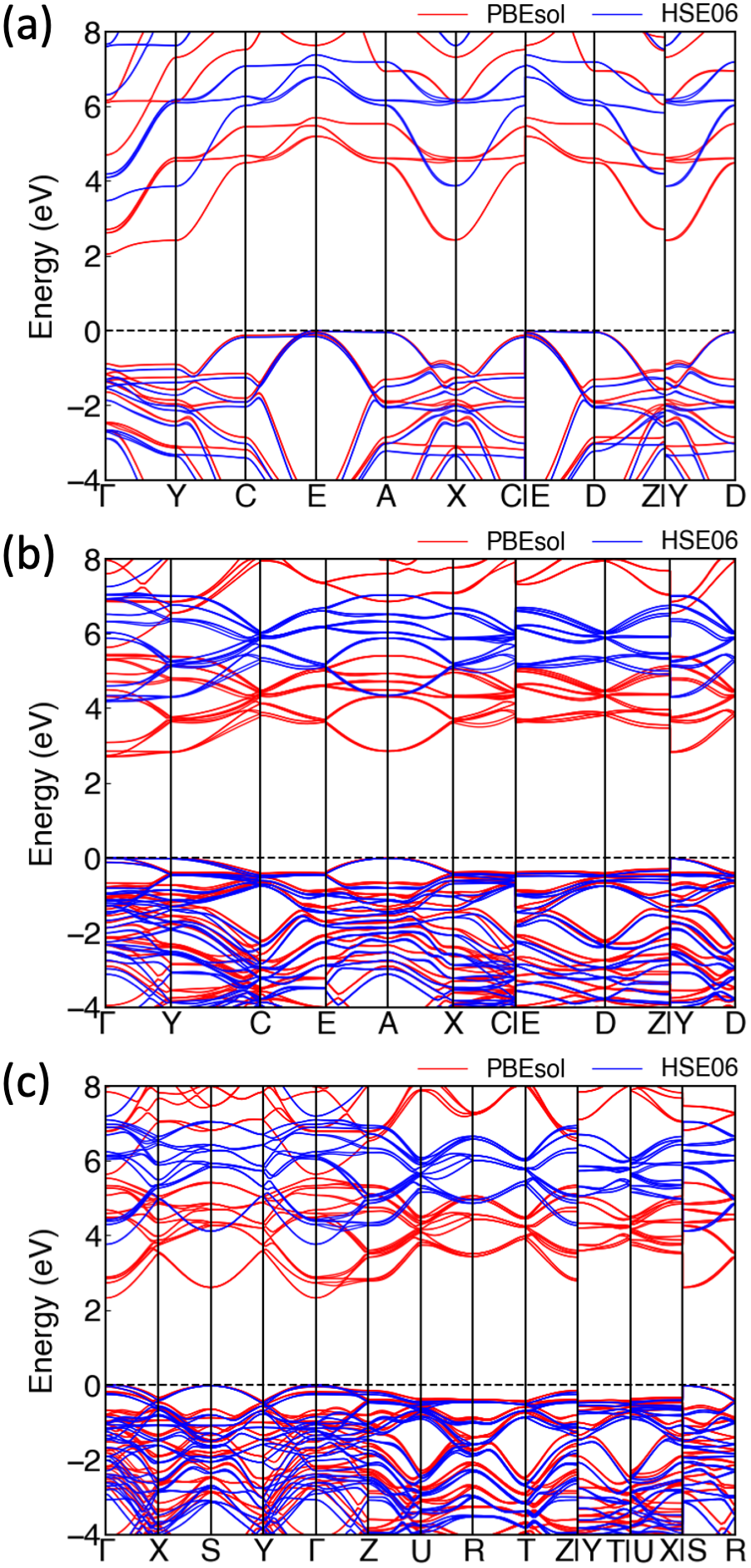}
\caption{\label{fig:band} Electronic band structure of (a) $Pm$, (b) $Cc$ and (c) $Pmc2_{1}$ phases. The valence band maximum is placed at 0 eV. Red and blue solid lines correspond to band states calculated with the exchange-correlation functional of PBEsol and HSE06, respectively.}
\end{figure}

Next, a comparison of computed and measured band gaps can help to further narrow down the potentially realized thin film phases of NaNbO$_{3}$. For the orthorhombic NaNbO$_{3}$/NdGaO$_{3}$ film, 
the band gap was previously determined experimentally to be approximately~3.8~eV\cite{BinAnooz:2022}.
For the monoclinic NaNbO$_{3}$/DyScO$_{3}$ film, an indirect band gap of 3.9-4.0~eV was measured, whereas the smallest observed direct optical transition corresponds to a gap of 4.72~eV\cite{BinAnooz:2022}.
To compare to these experiments, we computed the band structures of two monoclinic and five orthorhombic candidates at the HSE06 level of theory, as shown in Fig.~\ref{fig:band}~ for $Pm$, $Cc$, and $Pmc2_{1}$ phases.
For the orthorhombic phases ($Cmcm$, $Pca2_{1}$, $Pbcm$, $Pnma$, and $Pmc2_{1}$), we obtain band gaps of  3.01, 3.90, 3.74, 3.23, and 3.76~eV, respectively. Meanwhile, the monoclinic $Pm$ and $Cc$ phases exhibit band gaps of 3.46~eV and 4.17~eV.

\begin{figure}
\includegraphics[width=0.80\columnwidth]{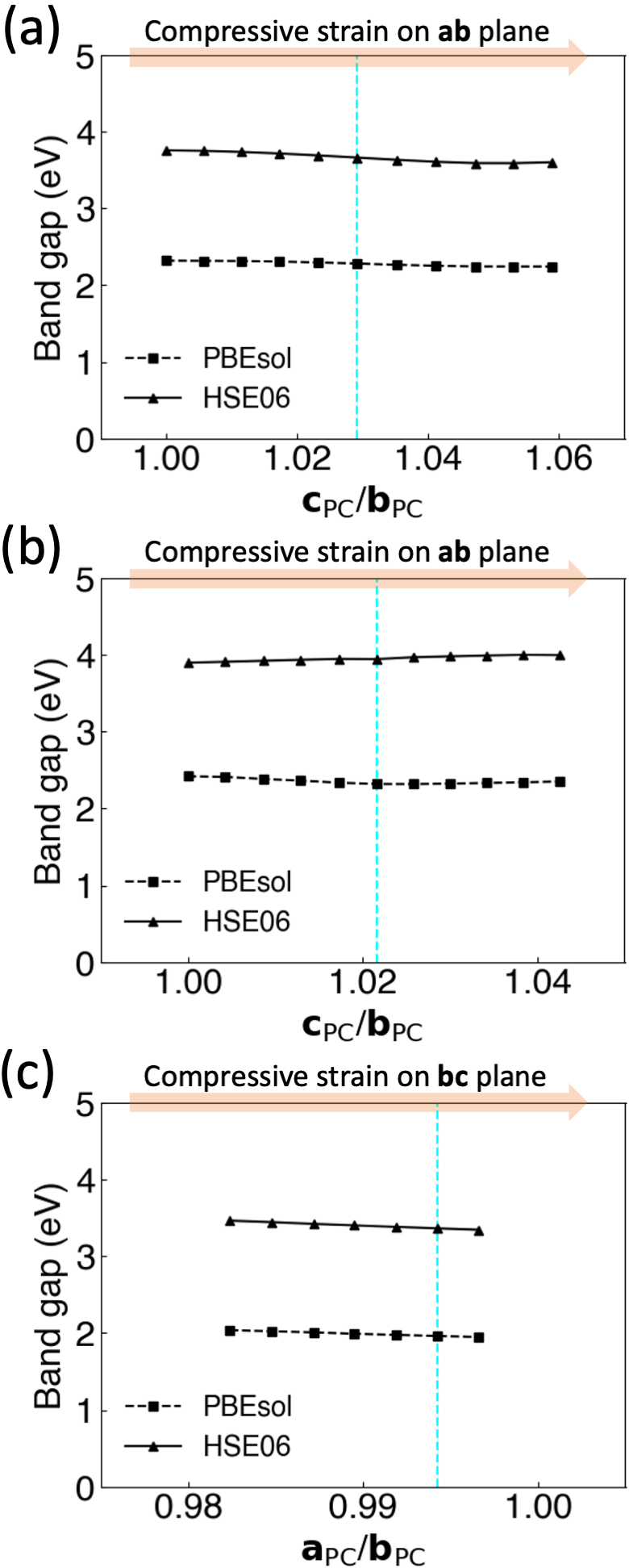}
\caption{\label{fig:gap} Band gap change of (a) $Pmc2_{1}$, (b) $Pca2_{1}$, and (c) $Pm$ in terms of the compressive strain strength. This compressive strain is applied on \textbf{ab} and \textbf{bc} plane for $Pmc2_{1}$ and $Pm$. The compressive strain is applied on \textbf{ab} for $Pmc2_{1}$ and $Pca2_{1}$ and on \textbf{bc} plane for $Pm$. The out-of-plane angle of $Pmc2_{1}$ and $Pca2_{1}$ is an angle of the electrical polarization vector away from $\mathbf{ab}$ plane. The in-plane angle of $Pm$ is an angle of the electrical polarization vector away from [010]$_{\mathrm{PC}}$. The cyan dashed lines correspond to the strain from the measurement. Dashed and solid lines correspond to the band gap results from calculations with the exchange-correlation functional of PBEsol and HSE06.}
\end{figure}

Utilizing a hybrid functional, here HSE06, is crucial in these calculations since it effectively mitigates the substantial underestimation of band gaps\cite{Krukau:2006} typically observed with semi-local functionals such as PBEsol. By this means, a (semi-)quantitative comparison to the experiment becomes possible. 
For instance, the calculated HSE06 band gaps of our bulk reference phases $Pm$-$3m$ and $Pbcm$, as depicted in Fig.~S1\cite{Supplement}, are 2.84~eV and 3.74~eV, which compares fairly well to the measured values\cite{Li:2012} of 3.29~eV and 3.45~eV. Still, there are some remaining discrepancies that can be attributed to the absence of electron-phonon coupling effects, which can substantially influence the band gap, especially for perovskite materials\cite{Wu:2020, Zacharias:20204h9}. Similarly, thin film effects and strain can affect the actual band gap, whereby our calculations suggest that the influence of the latter is largely negligible, as illustrated in Fig.~\ref{fig:gap}. 

Given the margin of uncertainty in the calculations and the fact that all computed band gaps are relatively close to the measured ones, it is hardly possible to reliably pinpoint the realized thin film solely on band-gap comparisons.
Still, a qualitative analysis can yield some additional insights.
Since the monoclinic sample exhibits an indirect gap in the experiment, the monoclinic $Cc$ phase, which possesses a direct band gap in the calculation, can be ruled out as a candidate. 
Conversely, the monoclinic $Pm$ phase displays an indirect band gap of 3.46~eV and a direct band gap of 4.49 eV at $\Gamma$ point in the calculations, which aligns well with the measured values.

In contrast, the measured electrical polarization in the NaNbO$_{3}$ films allows better insights into the actually realized polymorph.
To identify the phase of the NaNbO$_{3}$ film, we investigated the electrical polarization of all the orthorhombic and monoclinic phases listed.
Among the orthorhombic phases, $Pca2_{1}$ and $Pmc2_{1}$ show a spontaneous polarization of 0.20~C/m$^{2}$ along [100]$_{\mathrm{PC}}$ and 0.50~C/m$^{2}$ along [011]$_{\mathrm{PC}}$, respectively.
However, experimental observations for the NaNbO$_{3}$/NdGaO$_{3}$ sample indicate an out-of-plane polarization along [001]$_{\mathrm{PC}}$, which does not align with any of the calculated results.
Specifically, the calculated polarization of $Pca2_{1}$ directs in-plane [100]$_{\mathrm{PC}}$ perpendicular to the measured direction, while that of $Pmc2_{1}$ deviates by 45$^{\circ}$ from the measured direction.
The calculated polarization of the monoclinic $Pm$ phase is  0.58~C/m$^{2}$ along [011]$_{\mathrm{PC}}$ and thus closely matches the one observed in the NaNbO$_{3}$/DyScO$_{3}$ experimental sample.
Let us note that also the monoclinic $Cc$ phase would exhibit a compatible polarization of 0.58~C/m$^{2}$ along [111]$_{\mathrm{PC}}$; however, the $Cc$ phase had been ruled out before due to their
incompatible unitcell shape. To contextualize these numbers, we have further validated the predictive power of the polarization calculations for the $R3c$ phase, the only low-temperature 
ferroelectric bulk phase of NaNbO$_{3}$. Here, we obtain a spontaneous polarization of 0.57~C/m$^{2}$ along the [111]$_{\mathrm{PC}}$ direction, in excellent agreement with the experimental value 
of 0.57~C/m$^{2}$ obtained by Mishar \emph{et al.}~\cite{Mishra:2007}

\begin{figure}
\includegraphics[width=0.95\columnwidth]{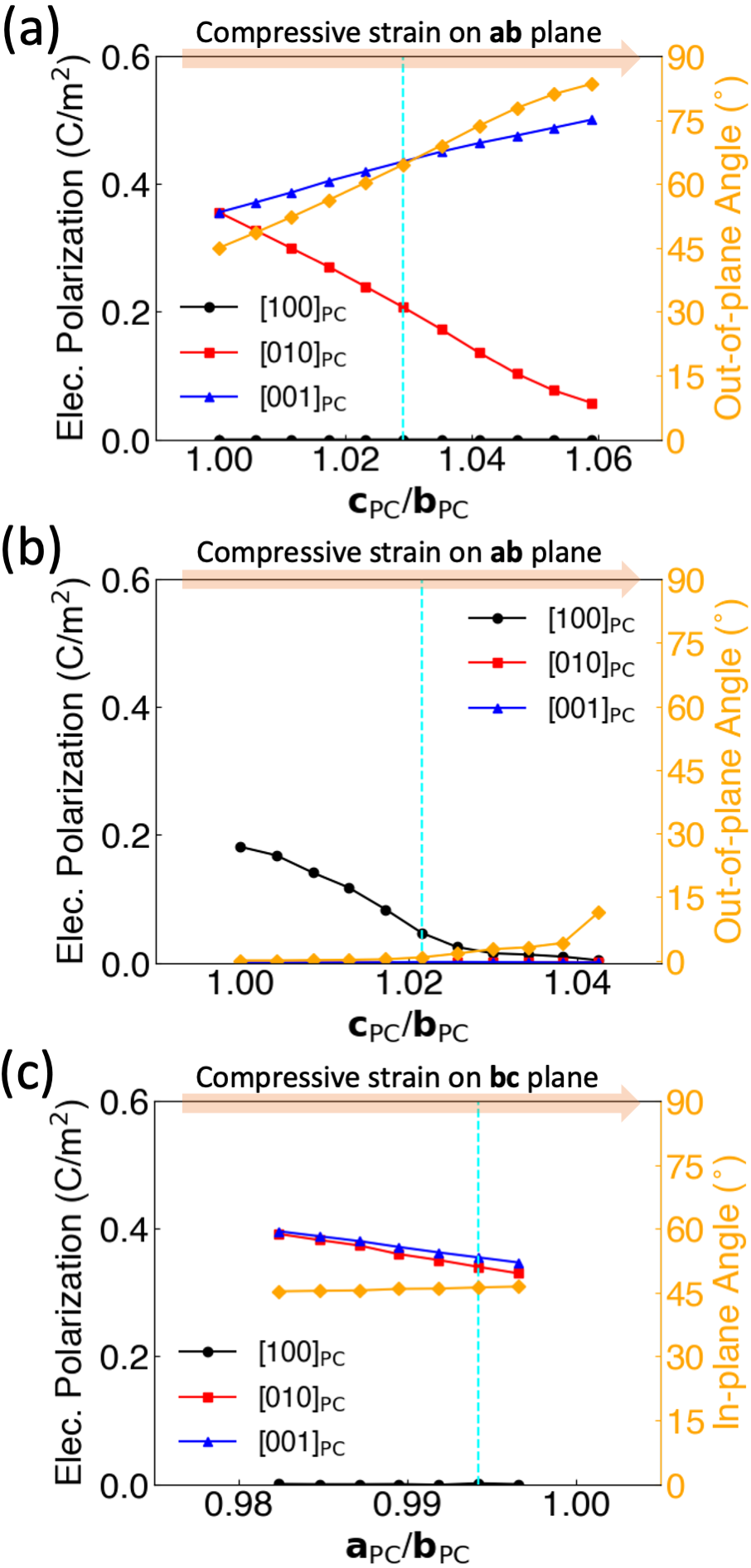}
\caption{\label{fig:strain} Electrical polarization and its directional angle change of (a) $Pmc2_{1}$, (b) $Pca2_{1}$ and (c) $Pm$ in terms of the compressive strain strength. The compressive strain is applied on \textbf{ab} for $Pmc2_{1}$ and $Pca2_{1}$ and applied on \textbf{bc} plane for $Pm$. The out-of-plane angle of $Pmc2_{1}$ and $Pca2_{1}$ is an angle of the electrical polarization vector away from $\mathbf{ab}$ plane. The in-plane angle of $Pm$ is an angle of the electrical polarization vector away from [010]$_{\mathrm{PC}}$. The cyan dashed lines correspond to the strain from the measurement\cite{Schmidbauer:2014, BinAnooz:2015}.}
\end{figure}

Despite the encouraging results obtained for the calculations of polarization in the bulk, the role of strain cannot be neglected.
To investigate this, we applied compressive strain on the $\mathbf{ab}$ plane of the relaxed orthorhombic $Pmc2_{1}$ structure by fixing the strained lattice parameters and monitoring the change in electrical polarization, as illustrated in Fig.~\ref{fig:strain} (a). When compressive strain is applied, the relaxation of the lattice parameter $\mathbf{c}_{\mathrm{PC}}$ induces changes in the lattice parameter ratio $\mathbf{c}_{\mathrm{PC}}$/$\mathbf{b}_{\mathrm{PC}}$ and affects the displacements of the Nb atoms.
In turn, the angle of the electrical polarization away from the $\mathbf{ab}$ plane increases from its bulk value of 45$^{\circ}$ and approaches  90$^{\circ}$ as the applied strain becomes stronger.
Qualitatively, this explains the alignment of the polarization observed in experiment~\cite{Schwarzkopf:2012, Schwarzkopf:2014}.
However, the thereto necessary strains slightly deviate from each other. In the experiment, a strong out-of-plane electrical polarization is observed already with a strain of about 1.03 $\mathbf{c}_{\mathrm{PC}}/\mathbf{b}_{\mathrm{PC}}$ lattice ratio, whereas strain larger than 1.06 $\mathbf{c}_{\mathrm{PC}}/\mathbf{b}_{\mathrm{PC}}$ lattice ratio is required for that in the calculations. We attribute this quantitative disagreement to the fact that only bulk phases and not thin films are modeled in our first-principles calculations.
In contrast, the spontaneous polarization of the orthorhombic phase $Pca2_{1}$ remains unchanged along [100]$_{\mathrm{PC}}$ under compressive strain, as shown in Fig.~\ref{fig:strain} (b).
For NaNbO$_{3}$/DyScO$_{3}$ case, the direction of electrical polarization in the monoclinic $Pm$ phase remains robust against compressive strain applied to the $\mathbf{bc}$ plane and continues to align with [011]$_{\mathrm{PC}}$, regardless of the strain strength, as shown in Fig.~\ref{fig:strain} (c).
Therefore, the strained orthorhombic $Pmc2_{1}$ phase emerges as the primary candidate for the phase of the NaNbO$_{3}$ film on the NdGaO$_{3}$ substrate, exhibiting varying electrical polarization from [011]$_{\mathrm{PC}}$ to [001]$_{\mathrm{PC}}$ under the compressive strain.
The monoclinic $Pm$ phase is identified as the phase of the NaNbO$_{3}$ film under compressive strain on the DyScO$_{3}$ substrate, maintaining a robust polarization along [011]$_{\mathrm{PC}}$.

\section{\label{sec:conc}Conclusion}
We implemented an investigation using the first-principles method, combining previously obtained experimental clues\cite{Schwarzkopf:2012, Schwarzkopf:2014, Schmidbauer:2014, Sellmann:2014, BinAnooz:2015, BinAnooz:2022, Guimaraes:2022} to determine the structural phase of NaNbO$_{3}$ films on NdGaO$_{3}$ and DyScO$_{3}$ substrates.
Our study employed DFT study to examine ten different phases of bulk NaNbO$_{3}$ candidates, which encompassed cubic $Pm$-$3m$, tetragonal $P4/mbm$, trigonal $R3c$, two monoclinic $Cc$ and $Pm$, and five orthorhombic $Cmcm$, $Pca2_{1}$, $Pbcm$, $Pnma$, and $Pmc2_{1}$.
A quantitative comparison of lattice parameters allowed us to narrow down the possible crystal structures and revealed the presence of strain effects on the orthorhombic phase.
Additionally, we explored the electronic band gap energies of both monoclinic and orthorhombic phases.
However, this information alone was insufficient to pinpoint the exact phase of the NaNbO$_{3}$ film.
In contrast, electrical polarization offered more intuitive insights and ultimately led to the identification of the NaNbO$_{3}$ film phase.
Our findings suggested that the NaNbO$_{3}$/NdGaO$_{3}$ sample likely possessed a strained $Pmc2_{1}$ structure, with electrical polarization rotating toward [001]$_{\mathrm{PC}}$ from [011]$_{\mathrm{PC}}$ under the influence of strain.
Meanwhile, the NaNbO$_{3}$/DyScO$_{3}$ sample was determined to have a strained $Pm$ structure with a robust electrical polarization along [011]$_{\mathrm{PC}}$.
This work paves the way for a more comprehensive exploration of the complex strained NaNbO$_{3}$ system through collaborative efforts between experimental and computational approaches.
These insights may result in advancing piezoelectric applications in the near future.

\begin{acknowledgments}
This project was supported by the NOMAD Center of Excellence (European Union's Horizon 2020 research and innovation program, Grant Agreement No, 951786), the ERC Advanced Grant TEC1p (European Research Council, Grant Agreement No. 740233) and BigMax (the Max Planck Society's Research Network on Big-Data-Driven Magterials-Science).
This work was performed in the framework of GraFOx, a Leibniz-Science Campus partially funded by the Leibniz Association.
KK and CC are grateful to Prof. Dr. Matthias Scheffler (FHI) for his support and inspiring discussions, which greatly contributed to this work.
\end{acknowledgments}


\bibliography{apssamp}

\end{document}


\title[Spontaneous polarization in NaNbO$_{3}$ film on NdGaO$_{3}$ and DyScO$_{3}$ substrates]{Spontaneous polarization in NaNbO$_{3}$ film on NdGaO$_{3}$ and DyScO$_{3}$ substrates}
\author{K. Kang}
 \email{kang@fhi-berlin.mpg.de}
\affiliation{ 
The NOMAD Laboratory at the FHI of the Max-Planck-Gesellschaft and IRIS-Adlershof of the Humboldt-Universität zu Berlin, Faradayweg 4-6, 14195 Berlin, Germany
}

\author{S. Bin Anooz}
\affiliation{
Leibniz-Institut f\"{u}r Kristallz\"{u}chtung (IKZ), Max-Born-Str. 2, 12489 Berlin, Germany
}

\author{J. Schwarzkopf}
\affiliation{
Leibniz-Institut f\"{u}r Kristallz\"{u}chtung (IKZ), Max-Born-Str. 2, 12489 Berlin, Germany
}

\author{C. Carbogno}
\affiliation{%
The NOMAD Laboratory at the FHI of the Max-Planck-Gesellschaft and IRIS-Adlershof of the Humboldt-Universität zu Berlin, Faradayweg 4-6, 14195 Berlin, Germany
}

\maketitle

\begin{sidewaystable}
\caption{\label{tab:struc}Lattice parameters of cubic, tetragonal, and rhombohedral phases of bulk NaNbO$_{3}$ from the experiments (Exp.) and DFT-PBEsol calculations. The difference between the measured and calculated lattices is shown in percentage.}
\begin{ruledtabular}
\begin{tabular}{cccccccccccccccc}
 & \multicolumn{5}{c}{Cubic} & \multicolumn{5}{c}{Tetragonal} & \multicolumn{5}{c}{Rhombohedral} \\
 \cline{2-6}\cline{7-11}\cline{12-16}
 & \multicolumn{5}{c}{$Pm$-$3m$} & \multicolumn{5}{c}{$P4/mbm$} & \multicolumn{5}{c}{$R3c$} \\
 \cline{2-6}\cline{7-11}\cline{12-16}
Lattice & Exp.\cite{Villars:2016} & PBEsol & Diff. & HSE06 & Diff. & Exp.\cite{Villars:2016} & PBEsol & Diff. & HSE06 & Diff. & Exp.\cite{Villars:2016} & PBEsol & Diff. & HSE06 & Diff. \\
\hline
$\mathbf{a}_{\mathrm{PC}}$ & 3.953~\AA & 3.943~\AA & -0.2~\% & 3.937~\AA & -0.4~\% & 3.944~\AA & 3.871~\AA & -1.9~\% & 3.878~\AA & -1.7~\% & 3.901~\AA & 3.906~\AA & 0.1~\% & 3.963~\AA & -0.1~\% \\
$\mathbf{b}_{\mathrm{PC}}$ & 3.953~\AA & 3.943~\AA & -0.2~\% & 3.937~\AA & -0.4~\% & 3.944~\AA & 3.871~\AA & -1.9~\% & 3.878~\AA & --1.7~\% & 3.901~\AA & 3.906~\AA & 0.1~\% & 3.963~\AA & -0.1~\% \\
$\mathbf{c}_{\mathrm{PC}}$ & 3.953~\AA & 3.943~\AA & -0.2~\% & 3.937~\AA & -0.4~\% & 3.952~\AA & 3.954~\AA & 0.1~\% & 3.946~\AA & -0.2~\% & 3.901~\AA & 3.906~\AA & 0.1~\% & 3.963~\AA & -0.1~\% \\
$\mathbf{\alpha}_{\mathrm{PC}}$ & 90$^{\circ}$ & 90$^{\circ}$ & 0.0~\% & 90$^{\circ}$ & 0.0~\% & 90$^{\circ}$ & 90$^{\circ}$ & -0.1~\% & 90$^{\circ}$ & 0.0~\% & 90$^{\circ}$ & 89.2$^{\circ}$ & -0.9~\% & 89.1$^{\circ}$ & -1.0~\% \\
$\mathbf{\beta}_{\mathrm{PC}}$ & 90$^{\circ}$ & 90$^{\circ}$ & 0.0~\% & 90$^{\circ}$ & 0.0~\% & 90$^{\circ}$ & 90$^{\circ}$ & -0.1~\% & 90$^{\circ}$ & 0.0~\% & 90$^{\circ}$ & 89.2$^{\circ}$ & -0.9~\% & 89.1$^{\circ}$ & -1.0~\% \\
$\mathbf{\gamma}_{\mathrm{PC}}$ & 90$^{\circ}$ & 90$^{\circ}$ & 0.0~\% & 90$^{\circ}$ & 0.0~\% & 90$^{\circ}$ & 90$^{\circ}$ & -0.1~\% & 90$^{\circ}$ & 0.0~\% & 90$^{\circ}$ & 89.2$^{\circ}$ & -0.9~\% & 89.1$^{\circ}$ & -1.0~\% \\
\end{tabular}
\end{ruledtabular}
\end{sidewaystable}

\begin{sidewaystable}
\caption{\label{tab:struc}Lattice parameter of orthorhombic and monoclinic phases of bulk NaNbO$_{3}$ from the experiments (Exp.) and DFT-PBEsol calculations. The difference between the measured and calculated lattices is shown in percentage.}
\begin{ruledtabular}
\begin{tabular}{ccccccccccccccccc}
 & \multicolumn{14}{c}{Orthorhombic} & \multicolumn{2}{c}{Monoclinic} \\
 \cline{2-15}\cline{16-17}
 & \multicolumn{5}{c}{$Cmcm$} & \multicolumn{5}{c}{$Pbcm$} & \multicolumn{2}{c}{$Pca2_{1}$} & \multicolumn{2}{c}{$Pnma$} & \multicolumn{2}{c}{$Cc$} \\
 \cline{2-6}\cline{7-11}\cline{12-13}\cline{14-15}\cline{16-17}
Lattice & Exp.\cite{Villars:2016} & PBEsol & Diff. & HSE06 & Diff. & Exp. & PBEsol & Diff. & HSE06 & Diff. & PBEsol & HSE06 & PBEsol & HSE06 & PBEsol & HSE06 \\
\hline
$\mathbf{a}_{\mathrm{PC}}$ & 3.937~\AA & 3.858~\AA & -2.0~\% & 3.866~\AA & -1.8~\% & 3.881~\AA & 3.866~\AA & -0.4~\% & 3.864~\AA & -0.4~\% & 3.883~\AA & 3.895~\AA & 3.886~\AA & 3.889~\AA & 3.906~\AA & 3.913~\AA \\
$\mathbf{b}_{\mathrm{PC}}$ & 3.942~\AA & 3.903~\AA & -1.0~\% & 3.903~\AA & -1.0~\% & 3.915~\AA & 3.910~\AA & -0.1~\% & 3.921~\AA & 0.1~\% & 3.903~\AA & 3.908~\AA & 3.881~\AA & 3.885~\AA & 3.906~\AA & 3.913~\AA \\
$\mathbf{c}_{\mathrm{PC}}$ & 3.945~\AA & 3.905~\AA & -1.0~\% & 3.905~\AA & -1.0~\% & 3.915~\AA & 3.910~\AA & -0.1~\% & 3.921~\AA & 0.1~\% & 3.903~\AA & 3.908~\AA & 3.881~\AA & 3.885~\AA & 3.906~\AA & 3.913~\AA \\
$\mathbf{\alpha}_{\mathrm{PC}}$ & 90$^{\circ}$ & 90$^{\circ}$ & 0.0~\% & 90$^{\circ}$ & 0.0~\% & 89.4$^{\circ}$ & 89.3$^{\circ}$ & -0.1~\% & 89.3$^{\circ}$ & -0.2~\% & 89.3$^{\circ}$ & 89.3$^{\circ}$ & 89.3$^{\circ}$ & 89.5$^{\circ}$ & 89.2$^{\circ}$ & 89.2$^{\circ}$ \\
$\mathbf{\beta}_{\mathrm{PC}}$ & 90$^{\circ}$ & 90$^{\circ}$ & 0.0~\% & 90$^{\circ}$ & 0.0~\% & 90$^{\circ}$ & 90$^{\circ}$ & 0.0~\% & 90$^{\circ}$ & 0.0~\% & 90$^{\circ}$ & 90$^{\circ}$ & 90$^{\circ}$ & 90$^{\circ}$ & 89.2$^{\circ}$ & 89.2$^{\circ}$ \\
$\mathbf{\gamma}_{\mathrm{PC}}$ & 90$^{\circ}$ & 90$^{\circ}$ & 0.0~\% & 90$^{\circ}$ & 0.0~\% & 90$^{\circ}$ & 90$^{\circ}$ & 0.0~\% & 90$^{\circ}$ & 0.0~\% & 90$^{\circ}$ & 90$^{\circ}$ & 90$^{\circ}$ & 90$^{\circ}$ & 89.2$^{\circ}$ & 89.2$^{\circ}$ \\
\end{tabular}
\end{ruledtabular}
\end{sidewaystable}

\begin{figure}
\includegraphics[width=0.75\columnwidth]{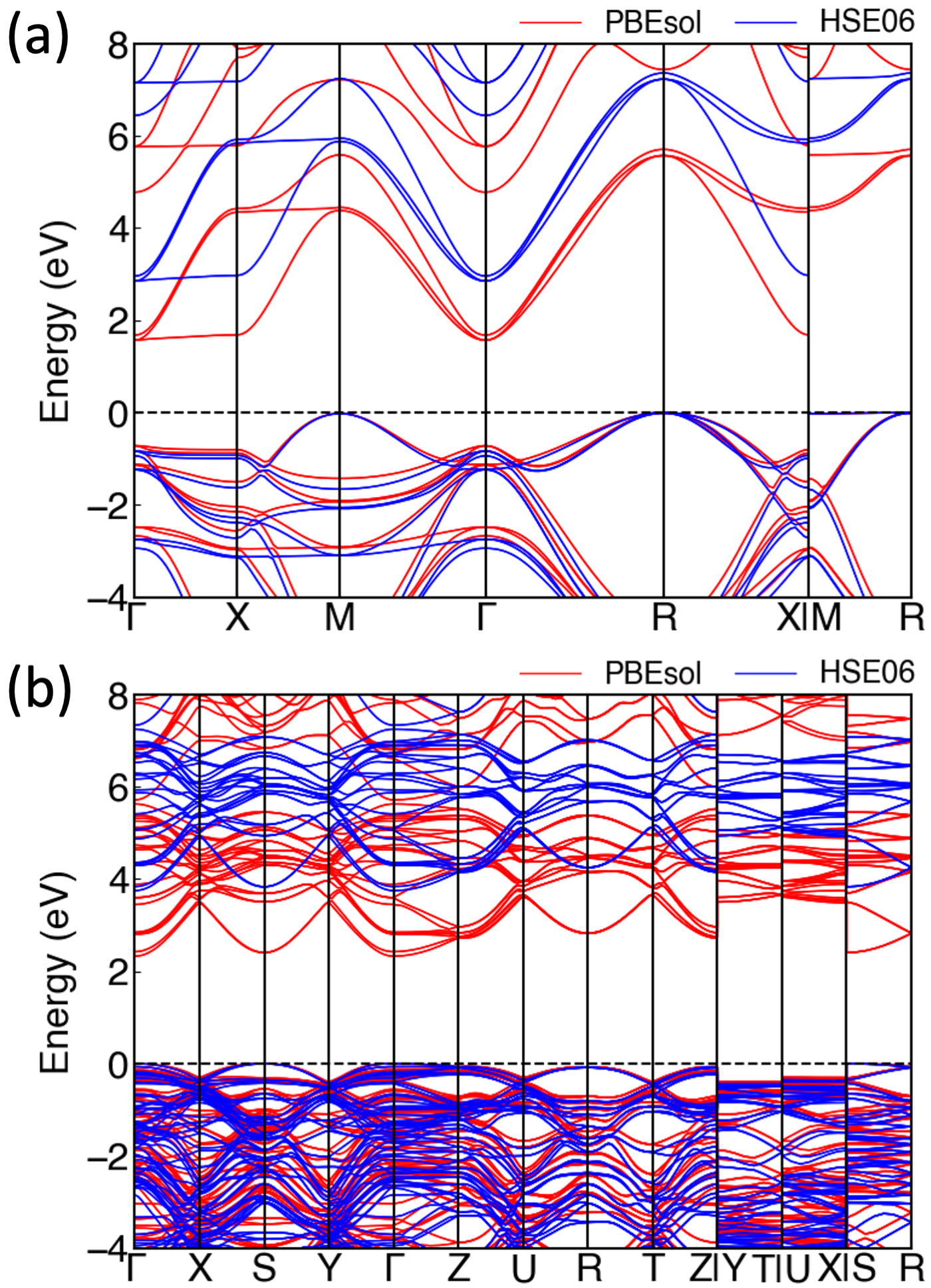}
\caption{\label{fig:band_supl2} Electronic band structures of (a) $Pm$-$3m$ and (b) $Pbcm$ phases. The valence band maximum is placed at 0 eV. Red and blue solid lines correspond to band states calculated with the exchange-correlation functional of PBEsol and HSE06, respectively.}
\end{figure}

\bibliographystyle{apsrev}
\bibliography{./supplementary.bib}